\documentclass[a4paper,twocolumn]{article}
\usepackage[DIV=15,BCOR=2mm,headinclude=true,footinclude=false]{typearea}
\usepackage{amssymb}
\usepackage{mathptmx}
\usepackage{amsmath}
\usepackage{caption}
\usepackage{subcaption}
\usepackage{graphicx}
\usepackage{lineno}
\usepackage{xurl}

\usepackage[breaklinks=true,linktocpage,breaklinks]{hyperref}
\usepackage{breakcites}
\usepackage{authblk}
\author[1,2,*]{Karsten Schuhmann}
\author[1,2]{Klaus Kirch}
\author[2]{Andreas Knecht}
\author[2]{Miroslaw Marszalek}
\author[3]{Francois Nez}
\author[2]{Jonas Nuber}
\author[4]{Randolf Pohl}
\author[1]{Ivo Schulthess}
\author[2]{Laura Sinkunaite}
\author[1]{Manuel Zeyen}
\author[1,2]{Aldo Antognini}

\affil[1]{Institute for Particle Physics and Astrophysics, ETH,  8093 Zurich, Switzerland}
\affil[2]{Paul Scherrer Institute, 5232 Villigen PSI, Switzerland}
\affil[3]{Laboratoire Kastler Brossel, UPMC-Sorbonne Universit\'es, CNRS,
ENS-PSL Research University, Coll\`ege de France, 75005 Paris, France.}
\affil[4]{Johannes Gutenberg-Universit\"at Mainz, QUANTUM, Institut f\"ur Physik \& Exzellenzcluster PRISMA, 55128 Mainz, Germany}
\affil[*]{Corresponding author: skarsten@phys.ethz.ch}

\title{Passive alignment stability and auto-alignment of multipass amplifiers based on Fourier transforms}

\begin{document}
\maketitle

\begin{abstract}
This study investigates the stability to tilts (misalignments) of Fourier-based multi-pass amplifiers, i.e., amplifiers where a Fourier transform is used to transport the beam from pass to pass.
%Multipass amplifiers based on optical Fourier transforms (Fourier-based amplifiers) exhibit cancellation of the thermal lens of the active medium.
Here, the stability properties of these amplifiers to misalignments (tilts) of their optical components has been investigated.
For this purpose, a method to quantify the sensitivity to tilts based on the amplifier small-signal gain has been elaborated and compared with measurements. 
To improve on the tilt stability by more than an order of magnitude a simple auto-alignment system has been proposed and tested. 
This study, combined with other investigations devoted to the stability of the output beam to variations of aperture and thermal lens effects of the active medium, qualifies the Fourier-based amplifier for the high-energy and the high-power sector.
\end{abstract}

\section{Introduction}

Multipass laser amplifiers are used to boost the output energy of laser oscillators owing to their smaller losses and higher damage thresholds~\cite{Chvykov:16, Keppler2012, Negel:15, Tokita:07}.
Because of their apparent simplicity, typically less attention is devoted to the design of multipass amplifiers compared to oscillators~\cite{Kogelnik:65, siegman1986lasers, Magni1986}.

Multipass amplifiers are commonly based on relay imaging (4f-imaging) from active medium to active medium as the imaging  provides  identical beam size for each passes at the active medium. 
The propagation in the 4f-based amplifier takes thus the simple form \cite{Simmons:78}
\begin{equation*}
AM-4f-AM-4f-AM-4f....\,\,,
\end{equation*}where AM indicates a pass on the active medium, and 4f a relay imaging.
However, in this design the changes of the phase front curvature caused by the thermal lens at the active medium are adding up from pass to pass, resulting  in output beam characteristics (size, divergence, quality) strongly dependent on the thermal lens of the active medium.

In this study,  Fourier-based multipass amplifiers are considered, i.e., multipass amplifiers where the propagation from active medium to active medium is accomplished using an optical Fourier transform.
The beam propagation in the Fourier-based amplifiers takes the form \cite{Schuhmann:18a}
\begin{align*}
AM& - Fourier - AM - SP - AM - Fourier \\
 - AM& - SP - AM - Fourier - AM - SP - AM...\,\,,
\end{align*}
where  SP represents a short free propagation and Fourier any propagation that performs an optical Fourier transform.
In contrast to the 4f-based amplifiers,   Fourier-based amplifiers show  output beam characteristics which are independent on variations of the thermal lens ensued by the pumped active medium~\cite{Schuhmann:18a}.
This relies on the peculiar property of the optical Fourier transform that inverts the phase front curvature, so that the phase front distortions occurring in one pass  at the active medium are canceled (in first order) by the successive pass \cite{Schuhmann:18a}.
Another advantage of the Fourier-based  design is the transverse mode cleaning that occurs from the interplay between Fourier transform and soft aperture effects at the active medium, favoring TEM00-mode operation.
Therefore, the Fourier-based design represents an alternative to 4f-based designs especially at the energy and power frontiers where the thermal lens becomes one of the paramount limitations.

The  concept of Fourier-based amplifiers  has been introduced in~\cite{Antognini2009}
and  successively elaborated in~\cite{Schuhmann2015,Schuhmann:16,Schuhmann2017,Schuhmann:PHD,Schuhmann:18a}, where a detailed comparison between Fourier-based and state-of-the-art multipass amplifiers based on the 4f-relay imaging is presented. 
In contrast to these earlier considerations, here another crucial aspect of laser design is considered, namely, the stability of the amplifier performance to misalignments (tilts).

%n ~\cite{Schuhmann:18a} it has been demonstrated that Fourier-based amplifiers show self-compensation of the thermal lens and aperture effects of the active medium. 
%
%\color{red} A short introduction to this type of laser amplifier is given in Section \ref{Realization}. \color{black} 
%
%Moreover, the interplay of aperture effects and the Fourier transform leads to an effective mode filtering that results in efficient laser operation in the TEM00 mode.
%
%On the contrary, the 4f-based design leads to an accumulation of the optical-phase-delay that occurs at the active medium, resulting in output beam characteristics (size, divergence, quality) strongly dependent on thermal lens, aperture effects and beam distortions ensued by the pumped active medium.
%

This study reveals that Fourier-based designs have also excellent (passive) stability properties for misalignments (tilts).
Furthermore, it demonstrates theoretically and practically, that the Fourier-based design is well apt for a simple auto-alignment system that mitigates the beam excursion from the optical axis (perfect alignment) caused by the tilts of its various optical components.
Hence, this study, combined with the investigations of~\cite{Schuhmann:18a}, displays the excellent performance of the Fourier-based architecture with respect to thermal lens variations, aperture effects, beam quality, efficiency and alignment stability, qualifying it for high-power and high-energy applications.

\section{Method to evaluate the misalignment sensitivity}
\label{sec:method}

The following procedure is undertaken to evaluate the sensitivity of the considered multipass amplifiers to misalignments.
First, the size of the fundamental Gaussian beam (TEM00 mode) propagating in the amplifier is evaluated with the ABCD-matrix formalism applied to the complex Gaussian beam parameter $q$.
A complex ABCD-matrix is used to model the aperture effect occurring in the active medium \cite{siegman1986lasers,Schuhmann:PHD,Schuhmann:18a}.
As a second step, the excursion of the Gaussian beam from the optical axis (the beam propagates on the optical axis for perfect alignment) is computed making use of the ABCD-matrix formalism applied to the geometrical ray describing the axis of the Gaussian beam.
For this purpose, another ABCD-matrix  must be defined, describing the aperture effects to this ray propagation.
The excursion of the laser beam from the optical axis can be used as a measure of the sensitivity of the multipass system to misalignment effects.
Knowledge of this excursion can be exploited also to compute the decrease of the multipass amplifier gain caused by the misalignments.
The dependency of the gain (transmission) on tilts constitutes the sensitivity of the multipass amplifier to mirror tilts.

\begin{figure}[htbp]
\centering
\includegraphics[width= \linewidth]{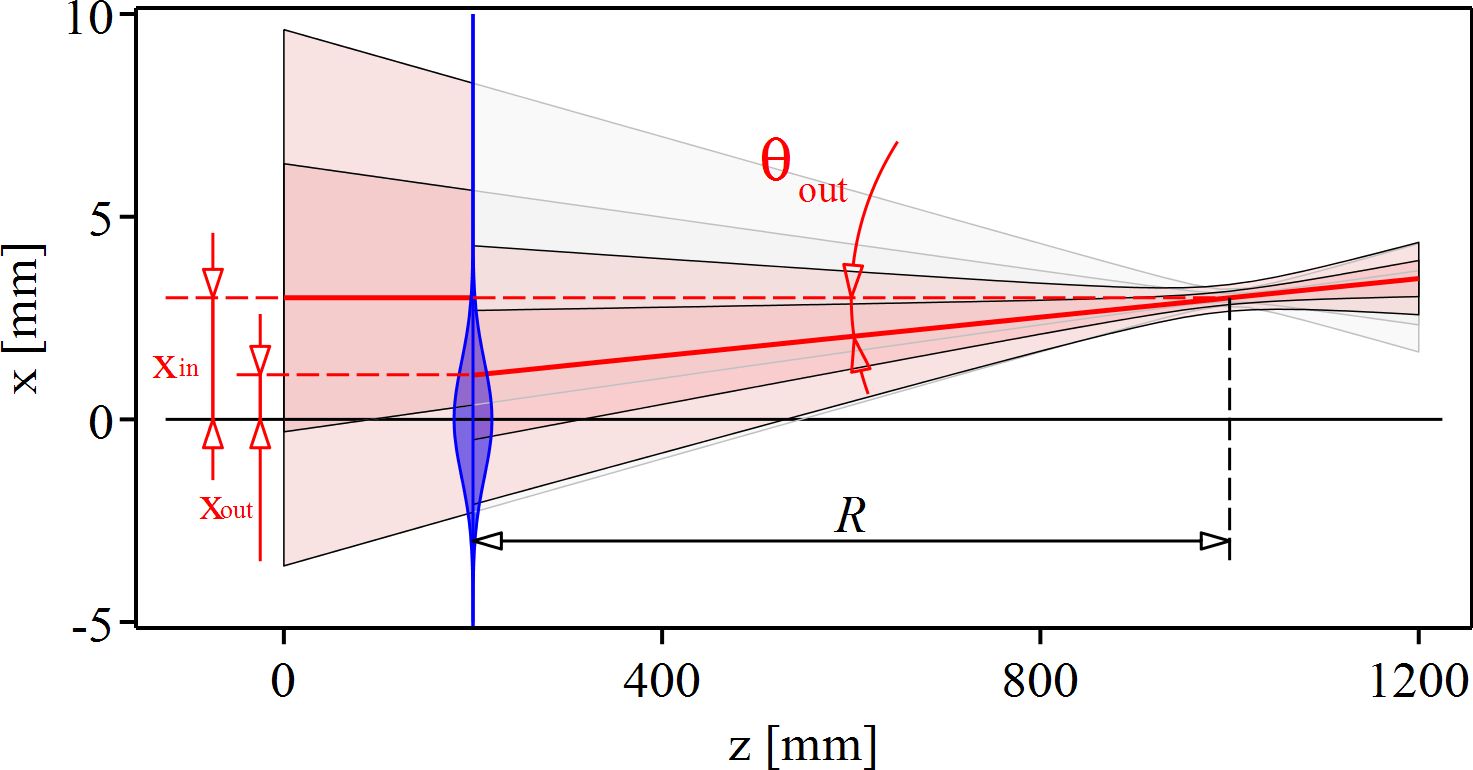}
\caption{Scheme showing the effect of a soft aperture on the beam propagation.  
The aperture whose transmission function is depicted by the blue area decreases the beam excursion from $ x_\mathrm{in}$ at the input plane to $x_\mathrm{out}$ at the output plane because it selects only part of the beam. 
Moreover, a non-vanishing  excursion of the beam at the impact plane ($x_\mathrm{in}\ne 0$) results in a change of the output beam angle $\theta_\mathrm{out} $.
Geometrical considerations can be used to deduce that  $\theta_\mathrm{out} - \theta_\mathrm{in}=(x_\mathrm{out} - x_\mathrm{in})/R$.
In this scheme it is assumed that the input beam is moving parallel to the optical axis, i.e. $\theta_\mathrm{in}=0$.  
Yet, the aperture does not affect the position of the beam focus.
The darker and lighter shaded red areas indicate the $\pm w$ and the $\pm 2w$ size of the beam.  
}
\label{fig:aperture-angle}
\end{figure}

In this study, it is assumed that losses in the multipass system occur only at the active medium whose position-dependent gain
(and absorption in the unpumped region) can be approximated by an average gain and a position-dependent transmission function (soft
aperture)~\cite{Schuhmann:18a}.
It is further assumed that the soft aperture of the active medium can be fairly approximated by a Gaussian aperture~\cite{Kogelnik:65, siegman1986lasers, Schuhmann:18a, Schuhmann:PHD}, i.e., an aperture with Gaussian intensity transmission function $\tau$:
\begin{equation}
  \tau (x,y)=e^{-\frac{x^2+y^2}{W^2}} \; ,
\end{equation}
where $x$ and $y$ are the transverse distances from the optical axis and $W$ the radius where the intensity is decreased to $1/e^2$.
Such an aperture transforms an input Gaussian beam with $1/e^2$-radius $w_\mathrm{in}$,  excursion $x_\mathrm{in}$ and  angle $\theta_\mathrm{in}$ w.r.t.  the optical axis into an output Gaussian beam with $1/e^2$-radius $w_\mathrm{out}$, excursion $x_\mathrm{out}$ and angle $\theta_\mathrm{out}$ given by
\begin{eqnarray}
  1/w^2_\mathrm{out} & = & 1/w^2_\mathrm{in}+1/W^2 \; , \label{eq:aperture-decrease}\\
  x_\mathrm{out}     & = & x_\mathrm{in}\frac{W^2}{w^2_\mathrm{in}+W^2} \; , \label{eq:aperture-offset}\\
  \theta_\mathrm{out}& = & \theta_\mathrm{in} -x_\mathrm{in}\frac{w^2_\mathrm{in}}{w^2_\mathrm{in}+W^2}\frac{1}{R} \; , \label{eq:aperture-angle}
\end{eqnarray}
where $R$ is the phase front radius of the beam at the aperture position.
Note that the phase front radius $R$ remains unaltered when the beam passes the aperture:
\begin{equation}
  R_\mathrm{out} = R_\mathrm{in} \equiv R \; .
    \label{eq:phase-front}
\end{equation}
Figure~\ref{fig:aperture-angle} can be used to deduce these equations relating the beam parameters before and after the aperture.
The relations expressed in Eqs.~(\ref{eq:aperture-decrease})-(\ref{eq:phase-front}) can be captured into the ABCD-matrix formalism.
Two ABCD-matrices can be defined to describe the Gaussian aperture at the active medium: the first applies to the complex parameter $q$ defined as \cite{Kogelnik:65, siegman1986lasers}
\begin{equation}
  \frac{1}{q}=\frac{1}{R} - i \frac{\lambda}{\pi w^2} \ ,
  \label{eq:complex-beam-parameter}
\end{equation}
where $\lambda$ is the wavelength of the laser beam,  $w$ and $R$ the local $1/e^2$-radius and phase front radius of the beam, respectively.
The corresponding  ABCD-matrix  is obtained from Eqs.~(\ref{eq:aperture-decrease}) and ~(\ref{eq:phase-front}). It takes the form
\begin{eqnarray}
  M_\mathrm{aperture}^{q}=\left[
    \begin{array}{cc} 1 & 0 \\ 
      - i\frac{\lambda}{\pi W^2} & 1
    \end{array}
\right]\ .
\end{eqnarray}
$M_\mathrm{aperture}^{q}$ is used to compute the size ($w$) and phase front radius ($R$) evolution along the propagation using the relation \cite{Kogelnik:65,siegman1986lasers}
\begin{equation}
  q_\mathrm{out}=\frac{Aq_\mathrm{in}+B}{Cq_\mathrm{in}+D} =\frac{q_\mathrm{in}}{-i\frac{\lambda}{\pi W^2}q_\mathrm{in}+1}  \ , 
  \label{eq:complex-beam-evolution}
\end{equation}
where $q_\mathrm{in}$ and  $q_\mathrm{out}$ are the $q$-parameters before and after  the aperture, respectively.

The second ABCD-matrix applies to the geometrical propagation of the beam axis and is obtained building on Eqs.~(\ref{eq:aperture-offset}) and (\ref{eq:aperture-angle}):
\begin{eqnarray}
  M_\mathrm{aperture}^\mathrm{geometry}=
  \left[
    \begin{array}{cc}
      \frac{W^2}{w_{in}^2+W^2} & 0 \\
      -\frac{w_{in}^2}{w_{in}^2+W^2} \frac{1}{R} & 1
    \end{array} \right] .
\end{eqnarray}
$M_\mathrm{aperture}^\mathrm{geometry}$ is used to calculate the
excursion $x_\mathrm{out} $ and angle $\theta_\mathrm{out} $ after passing
the aperture:
\begin{eqnarray}
  \left[
    \begin{array}{c}
     x_\mathrm{out}   \\
     \theta_\mathrm{out}  
    \end{array}
    \right]
        =
   \left[
     \begin{array}{cc}
       \frac{W^2}{w_{in}^2+W^2} & 0 \\
       -\frac{w_{in}^2}{w_{in}^2+W^2} \frac{1}{R} & 1
     \end{array} \right]
  \left[
    \begin{array}{c}
     x_\mathrm{in}   \\
     \theta_\mathrm{in}  
    \end{array}
    \right]
  .
  \label{eq:ABCD-geometrical}
\end{eqnarray}
Similar equations are valid for the other transverse direction ($y$-direction).

As visible from Eq.~(\ref{eq:aperture-decrease}), the passage through an aperture reduces the beam size, thereby generating power and intensity losses.
For a Gaussian beam on the optical axis, the power transmission through the aperture is given by
\begin{eqnarray}
  T_\mathrm{aperture}^\mathrm{aligned}=\frac{W^2}{w_\mathrm{in}^2 + W^2} \; .
    \label{eq:transmission-1}
\end{eqnarray}
This transmission is further decreased when the beam has an offset from the optical axis of $x_\mathrm{in}$ and $y_\mathrm{in}$:

\begin{equation}
  T_\mathrm{aperture}^\mathrm{mis-aligned} = e^{-2\frac{x_\mathrm{in}^2 + y_\mathrm{in}^2}{w_\mathrm{in}^2+W^2}}~\frac{W^2}{w_\mathrm{in}^2 + W^2} \ .
  \label{eq:transmission-2}
\end{equation}

The power transmission $T_\mathrm{tot}$ through the amplifier  is obtained by multiplying the transmissions of each pass at  the active medium:
\begin{equation}
  T_\mathrm{tot} =\displaystyle\prod_{n=1}^{N} T_\mathrm{aperture}^\mathrm{mis-aligned} [n] \; , 
  \label{eq:transmission-3}
\end{equation}
with $N$ representing the total number of passes at the active medium (or media) and $T_\mathrm{aperture}^\mathrm{mis-aligned}[n]$ the transmission at the $n$-th pass which depends on the beam size $w_\mathrm{in}$, on the deviation from the optical axis $x_\mathrm{in}$ and $y_\mathrm{in}$, and on the size of the aperture $W$ at the $n$-th pass.

It is important to note that this formalism assumes Gaussian apertures and Gaussian beams.
As mentioned before a Gaussian aperture transforms a Gaussian beam into a Gaussian beam of different size. 
A non-Gaussian aperture, on the contrary, leads to excitation of higher-order transverse beam components. Yet, in the Fourier-based amplifier design presented here, the higher-order components produced in one pass are filtered out in the next pass by the aperture itself.  Hence, the TEM00 component dominates, validating the use of Gaussian beams. 
In this paper $W=4w_\mathrm{in}$ is assumed because
a Gaussian aperture with $W=4w_\mathrm{in}$ produces a similar reduction of the fundamental mode size and a similar reduction of  the fundamental mode transmission compared to a super-Gaussian  aperture fulfilling the relation $w_\mathrm{in} \approx 0.7R_p$  \cite{Peng:15,Schuhmann:PHD,Negel:13}, where $R_p$  is the radius of the super-Gaussian pump spot.
The latter relation  is a rule of thumb typically applied in  the thin-disk laser community as it  provides most  efficient laser operation in the fundamental mode.
The assumption $W \approx 4w_\mathrm{in}$ when $w_\mathrm{in} \approx 0.7R_p$ is confirmed in Sec. \ref{Misalignment sensitivity} by the measurement of the small signal gain versus tilt angles.

\section{Realization of a multi-pass amplifier based on Fourier transform propagations}
\label{Realization}

\begin{figure}[tbp]
\centering
\includegraphics[width=  0.9 \linewidth]{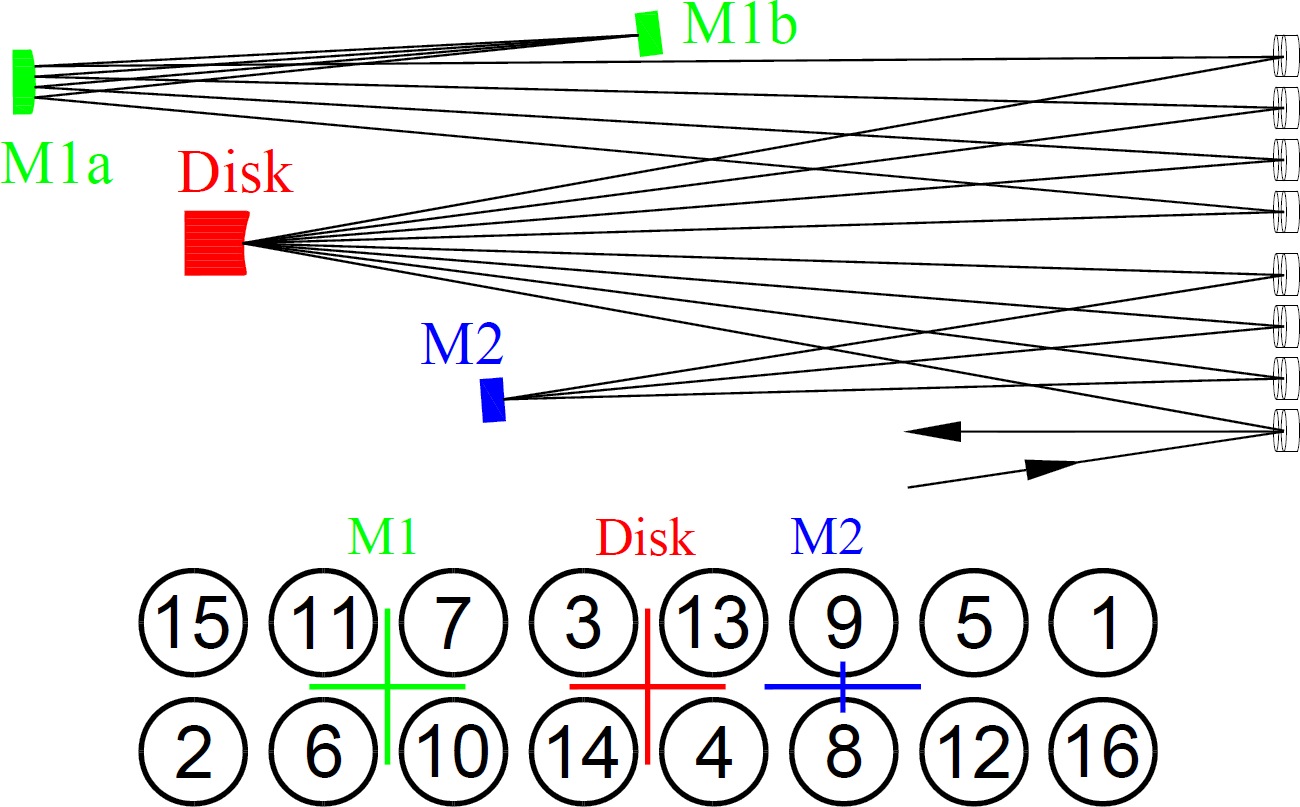}
\caption{(Top) Scheme of the realized Fourier-based amplifier with the corresponding beam path. 
The beam routing in the multipass amplifier is sustained by an array of flat mirrors.  (Bottom) Front view on the mirror array and its working principle. 
The mirrors \#1 to \#16 are numbered according to the sequence the beam passes them.
The beam enters the amplifier over array mirror \#1 and is reflected by the disk to \#2, from there over M1 (M1a-M1b-M1a) to \#3, over the disk to \#4, over M2 to \#5, and so forth until the disk is passed 8 times.
The crosses indicate the position of M1, M2 and disk ``projected'' on the mirror array.
They act as point reflectors in the mirror array plane for the propagation from mirror to mirror. 
Hereby we denote as mirror M1 the two-mirror system given in green  composed of the convex mirror M1a and the ``end-mirror'' M1b. }
\label{fig:scheme-old-amplifier}
\end{figure}

The investigations of alignment stability of Fourier-based amplifiers presented in this study are based without loss of generality on the specific thin-disk multi-pass amplifier shown in Fig.~\ref{fig:scheme-old-amplifier}.
The working principle of thin-disk lasers~\cite{Brauch:95, Giesen1994, Giesen2005,Giesen2007, Giesen2007b} is not presented in this work as it is of minor relevance for the argumentation exposed.
\begin{figure}[tbp]
\centering
\includegraphics[width=.45 \linewidth, angle=0]{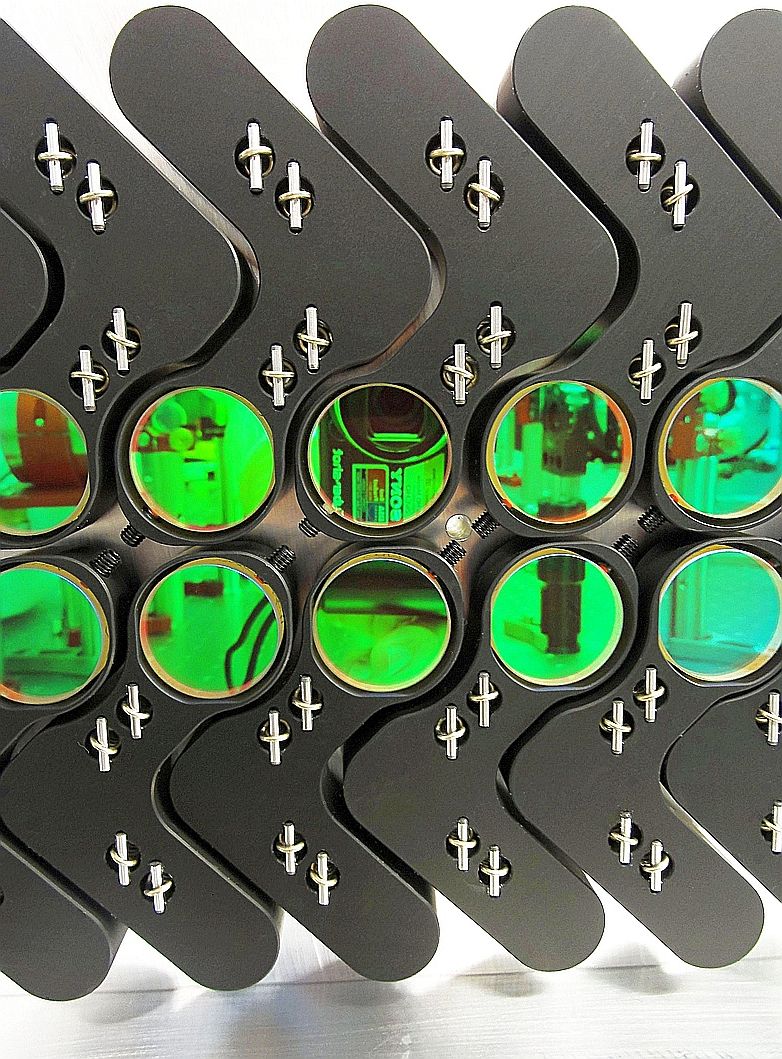} \hfill
\includegraphics[width=.3\linewidth]{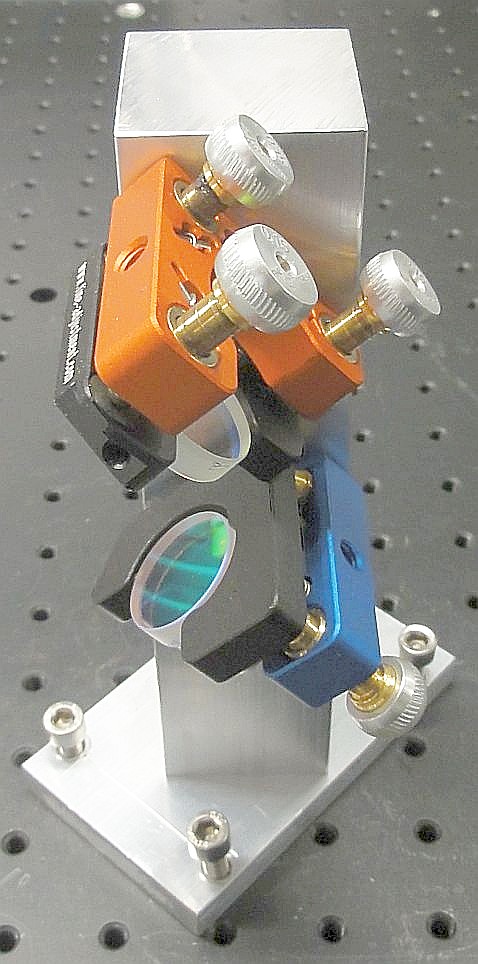}
\caption{(Right) Photo of the custom designed mirror array with L-shaped adjustable mirror holders.  
The horizontal spacing between the 1'' mirrors is 31~mm. 
(Left) Picture of the pair of $45^\circ$ mirrors acting as vertical retro-reflector that can assume the functionality of the mirror M2 in Fig.~\ref{fig:scheme-old-amplifier}. }
\label{fig:mirror-array} 
\end{figure}
In this amplifier, the beam is reflected and amplified 8 times (8-pass amplifier) on the disk while it is propagating between mirror M2 and mirror M1b.
The various passes having slightly different beam paths are realized using an array of flat mirrors whose working principle is shown in the bottom part of Fig.~\ref{fig:scheme-old-amplifier}.
L-shaped mirror holders were developed and placed as shown in Fig.~\ref{fig:mirror-array} to maximize tilt stability of the individual mirrors while minimizing the array size.
Commercially available mirror holders with similar alignment stability are significantly larger leading to substantially larger spacing between the mirrors.
The small size of the mirror array decreases the astigmatism related to incident angles and it guarantees that the various path lengths (especially between elements with non-vanishing dioptric power) are similar for all passes so that the beam size is reproduced from pass to pass.
For the same reason, the multi-pass propagation is designed to have same M1a, M1b, M2 and disk for all passes.
 In the design presented in this paper the disk and M1a have  a non-vanishing dioptric power while M1b, M2 and the array mirrors are flat.
This choice reduces the complexity and the costs and also simplifies the alignment procedure as only the radii of curvature of the mirror M1a and the distances M1a-disk and M1a-M1b must be adapted to realize the desired layout with the same beam size at each pass.

The back and forth propagation between the disk (AM) and M1 is described by an ABCD-matrix approximatively of the form
\begin{equation}
M_\mathrm{AM-M1-AM}\approx\left[ \begin{array}{l l}
    0 & B \\
     1/B & 0 \end{array} \right] \,
\label{eq:ABCD-matrix}
\end{equation}
which corresponds to the ABCD-matrix of a Fourier transform (see Ref.~\cite{Schuhmann:18a} for more details).
Differently, the back and forth propagation from the disk to M2, corresponds approximately to a short free propagation so that its ABCD-matrix takes the form
\begin{equation}
M_\mathrm{AM-M2-AM}\approx \left[ \begin{array}{l l}
    1 & L \\
     0 & 1 \end{array} \right] \ ,
\label{eq:ABCD-matrix-short}
\end{equation}
with a length $L$ short compared with the Rayleigh length of the beam, i.e.  with $L\ll \pi w_0^2/\lambda$, where $w_0^2$ is the beam waist.
Note that the beam size between the AM and mirror M2 is large and that the dioptric power of the mirror M2 is either zero or a minor correction to the free propagation.
Thus, the multipass amplifier depicted in Fig.~\ref{fig:scheme-old-amplifier}  follows the scheme
\begin{align*}
AM&-Fourier-AM-SP-AM-Fourier \\
-AM&-SP-AM-Fourier-AM-SP-AM-... \; .
\end{align*}
In this paper, the focus is on the stability properties of this Fourier-based multi-pass amplifier w.r.t. misalignments (tilts), especially of the active medium.
Two amplifier configurations based on the scheme of Fig.~\ref{fig:scheme-old-amplifier} are investigated: in the first configuration the M2 mirror is a flat back-reflecting mirror (acting almost as a ``end-mirror''), in the second configuration, the mirror M2 is replaced by a pair of mirrors oriented at 45$^\circ$ as shown in Fig.~\ref{fig:mirror-array} (b) that acts as a vertical (y-direction) retro-reflector.

\section{Sensitivity of a multipass amplifier to misalignment}
\label{Misalignment sensitivity}

\begin{figure}[htbp]
\centering
\includegraphics[width= \linewidth]{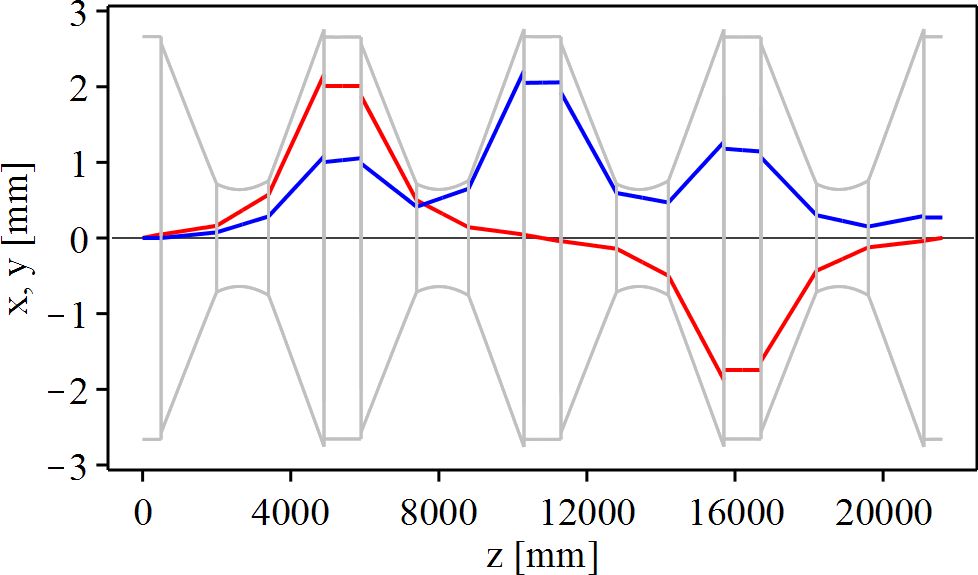}
\caption{
Excursion of the Gaussian beam axis w.r.t. the optical axis along the propagation in the Fourier-based amplifier of Fig.~\ref{fig:scheme-old-amplifier} computed using the ABCD-matrix formalism applied to ray optics and including aperture effects.  
The blue curve represents the beam excursion for a disk tilt of $\phi=50\;\mu$rad.  The red curve represents the beam excursion for an input beam with an angle of $100\;\mu$rad, w.r.t. a perfect alignment. 
As a comparison, the beam size evolution $\pm w$ along the propagation axis $z$ is shown by the grey curves.  
An aperture width of $W=10$~mm has been used in these plots.  
The vertical lines represent the positions of the disk (with a focusing dioptric power) and the positions of the defocusing mirror M1a, respectively. 
These plots apply for both the horizontal ($x$-) and the vertical ($y$-) directions.  }
\label{fig:tilt-propagation-1}
\end{figure}

The excursion of the laser beam from the optical axis while propagating in the multipass amplifier of Fig.~\ref{fig:scheme-old-amplifier} is shown in Fig.~\ref{fig:tilt-propagation-1}.
It has been calculated using the ABCD-matrix formalism applied to the geometric ray representing the Gaussian beam axis.
Aperture effects at the disk have been included via the ABCD-matrix of Eq.~(\ref{eq:ABCD-geometrical}) and are visible in Fig.~\ref{fig:tilt-propagation-1} by the sudden decrease of the beam size at the disk.
The excursion of the propagating beam from the optical axis has been evaluated for two different types of misalignment and compared to the evolution of the beam size along the optical axis.
In red is displayed the excursion for a tilt of the beam in-coupled into the multipass system, in blue for a tilt of the disk.
When neglecting aperture effects, at the 8$^\mathrm{th}$  pass there is a vanishing beam excursion for both types of misalignment.
Indeed, the multiplication of the four Fourier transform matrices that take place between the 1$^\mathrm{st}$ and the 8$^\mathrm{th}$ pass corresponds to the identity matrix.
Hence, disregarding the aperture effects and the short propagations of Eq.~(\ref{eq:ABCD-matrix-short}), the beam leaves the disk after the 8$^\mathrm{th}$ pass at exactly the same position and with the same angle w.r.t. the optical axis as in the 1$^\mathrm{st}$ pass.

In general, the beam cutoffs caused by the apertures at the active medium damp the beam excursions and angles relative to the optical axis.
For the special case of the 8$^\mathrm{th}$ pass (most stable pass with respect to thermal lens effects and thin-disk tilt), the beam angle is reduced compared to the 1$^\mathrm{st}$  pass, but the beam position  is slightly offset from the optical axis.
This small excursion decreases with increasing aperture size $W$.

Throughout this paper, using the rule of thumb diffused in the thin-disk laser community and justified in \cite{Schuhmann:PHD} a Gaussian aperture with $W=10$~mm is assumed for a pump diameter of 7~mm at FWHM, and a $1/e^2$ beam radius at the disk of $w=2.5$~mm.

\begin{figure}[tbp]
\centering
\includegraphics[width= \linewidth]{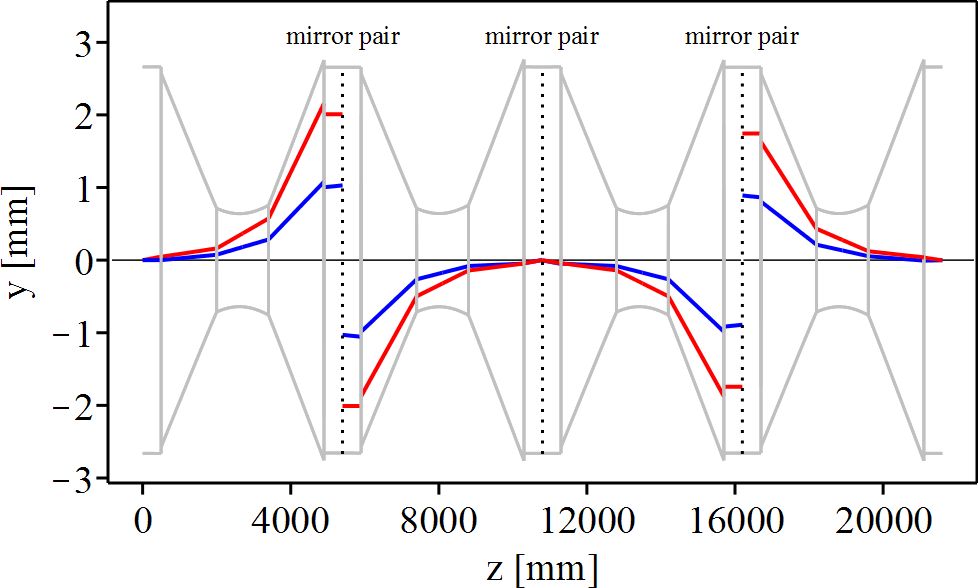}
\caption{Similar to Fig.~\ref{fig:tilt-propagation-1}, but for a multipass amplifier where M2 is replaced by a pair of $45^\circ$ mirrors acting as a vertical retro-reflector. 
%The position of the retro-reflector is indicated by the vertical dotted lines. 
The vertical dotted lines indicate the position of the retro-reflector. 
The excursion is shown only for the vertical component. 
The horizontal component is not affected by the vertical retro-reflector so that it follows the evolution shown in Fig.~\ref{fig:tilt-propagation-1}.  }
\label{fig:tilt-propagation-2}
\end{figure}

Inspired by the studies reported in \cite{Negel:15} also a modified version of the multipass amplifier shown in Fig.~\ref{fig:scheme-old-amplifier} has been investigated in which mirror M2 is replaced by a pair of $45^\circ$ mirrors acting as a vertical (y-direction) retro-reflector.
This mirror pair inverts the tilt and the excursion of the beam from the optical axis in the  y-direction while it does not affect the tilt and the excursion in the x-direction.
Hence, as already shown in \cite{Schuhmann:15}, a vertical retro-reflector significantly increases the alignment stability in the vertical direction.
Here, the evaluation of the alignment stability of \cite{Schuhmann:15} has been advanced to include aperture effects.
As can be seen by comparing Fig.~\ref{fig:tilt-propagation-2} to Fig.~\ref{fig:tilt-propagation-1}, this modification leads to a significant reduction of the beam excursions in particular for the $4^{\mathrm{th}}$ and the $5^{\mathrm{th}}$ pass but also to a reduction of the excursion and angle of the out-coupled beam (leaving the 8$^{\mathrm{th}}$ pass).

%It should be noted that the vertical retro-reflector causes an inversion of the beam angle and excursion in vertical direction, but it does not affect the beam properties in the horizontal direction.
%
Note that in the amplifier designs presented in this paper (and in the majority of the amplifier designs) the misalignment of the beam in the x- and the y-direction can be treated independently.
We chose to implement a retro-reflector in vertical direction to reduce %the amplifiers sensitivity to vertical misalignments.
 the misalignment effect caused by the  hot air wedge at the front side of the disk \cite{Negel:15,dietrich2017passive}, reducing the sensitivity to variations of the pump power.

The use of a corner-cube reflector with 3 mirrors arranged at an angle of $54^\circ$ and acting as retro-reflectors would provide inversion of the beam excursions and angles for both the horizontal and the vertical directions, resulting in an enhanced alignment stability in both directions.
\begin{figure}[bth]
\centering
\includegraphics[width=.5 \linewidth]{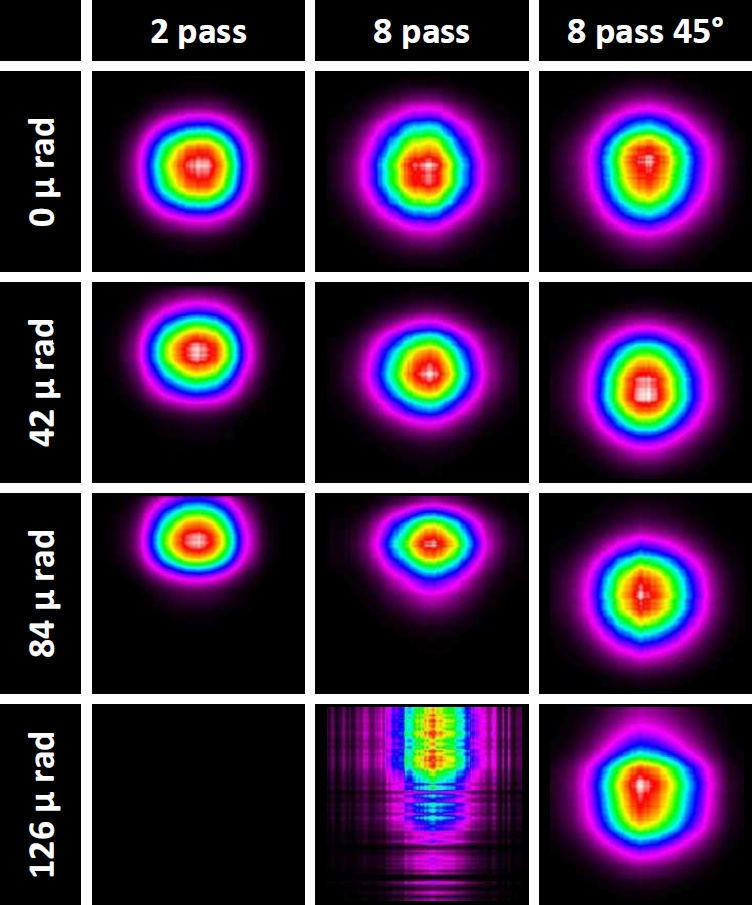}
\caption{
Measured beam profiles for various disk tilts for three multipass configurations based on  Fig.~\ref{fig:scheme-old-amplifier}.
The colors represent the beam intensity. 
Each picture is normalized to its intensity maximum.
These images show how beam excursions and distortions increase with misalignment.
Note that the first row represents  the reference point as it shows the beam position for the aligned amplifier.
The first column is the excursion at the $2^{\mathrm{nd}}$ pass, the second column at the $8^{\mathrm{th}}$ pass, and the third column also at the $8^{\mathrm{th}}$ pass but for an amplifier where M2 is replaced by a pair of $45^\circ$ mirrors acting as a vertical retro-reflector.  
The $\phi=126\;\mu$rad tilt causes such a large beam excursion already at the second pass that the beam is deviated outside the aperture of the beam profiler used to record the images. 
At the $8^{\mathrm{th}}$-pass the beam excursion as shown in Fig.~\ref{fig:tilt-propagation-1} is much smaller so the beam enters the profiler. 
Yet, this beam is fully distorted because of hard aperture effects occurring at the $4^{\mathrm{\mathrm{th}}}$ pass where the excursion is maximal. 
When the retro-reflector is introduced, the excursion at the $4^{\mathrm{th}}$ pass vanishes so that the beam distortions at the $8^{\mathrm{th}}$ pass disappear. 
This results in a beam at the $8^{\mathrm{th}}$ pass with almost a Gaussian profile and a small excursion from the unperturbed position given in the first row.}
\label{image8}
\end{figure}

The beam excursions simulated in Figs.~\ref{fig:tilt-propagation-1} and ~\ref{fig:tilt-propagation-2} have been confirmed by measurements summarized  in Fig.~\ref{image8}.
These measurements show beam profiles and positions (excursions) for various tilts of the disk for three configurations:
the first column at the $2^\mathrm{nd}$ pass of the Fourier-based multi-pass amplifier of Fig.~\ref{fig:scheme-old-amplifier}, the second column at the 8$^\mathrm{th}$ pass of the same amplifier, and the third column at the 8$^\mathrm{th}$ pass of the same amplifier but the mirror M2 replaced by a pair of $45^\circ$ mirrors acting as a vertical retro-reflector.
Note that when a beam is reflected at a tilted disk, it acquires an angle $\theta $ w.r.t. the optical axis corresponding to twice the disk tilt, i.e.  $\theta=2 \phi$, where $\phi$ is the tilt of the disk axis from perfect alignment.

As predicted by the simulations of Figs.~\ref{fig:tilt-propagation-1} and \ref{fig:tilt-propagation-2} the beam excursions from the unperturbed position (for $0~\mu$rad tilts) are larger at the $2^\mathrm{nd}$ pass than at the $8^\mathrm{th}$ pass.
With increasing tilt, the profile at the $8^\mathrm{th}$ pass (second column) is distorted.
These distortions visible for a tilt of $\phi=84~\mu$rad originate mainly at the $4^\mathrm{th}$ pass, where the beam excursions are maximal (see Fig.~\ref{fig:tilt-propagation-1}).
Indeed, for large excursions the aperture at the disk is poorly approximated by a Gaussian transmission function resulting in a transmitted beam with non-Gaussian (non-symmetric) profile.
Larger distortions and even fringe effects occurs for a tilt of $\phi=126~\mu$rad at the $8^\mathrm{th}$ pass (second column).
These fringes originate from hard apertures and beam cut-offs at the edge of optical components mostly at, or around the $4^\mathrm{th}$ pass, where beam excursions are maximal.

A comparison between second and third column discloses the improvements in term of stability to misalignment yielded by the use of the M2 retro-reflector.
It suppresses beam excursions at the $8^\mathrm{th}$ pass.
Even more, beam distortions originating around the $4^\mathrm{th}$ pass are suppressed as the retro-reflector strongly restrains the beam excursions at the intermediate passes (compare Fig.~\ref{fig:tilt-propagation-1} to Fig.~\ref{fig:tilt-propagation-2}).

\begin{figure}[tb]
\centering
\includegraphics[width= 0.8 \linewidth]{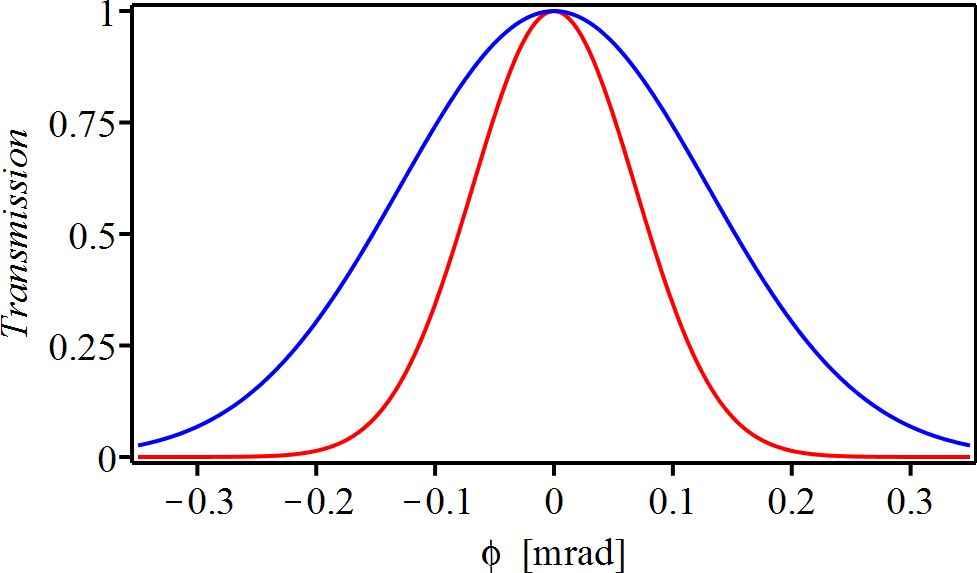}
\caption{The red curve represents the tilt-dependent part of $G_\mathrm{8-pass}^\mathrm{M2-single}$ ($e^{-1.07\cdot {10}^7 \phi
    ^2}$) for the 8-pass amplifier of Fig.~\ref{fig:scheme-old-amplifier} as a function of the disk tilt $\phi$. 
  It corresponds to the tilt-dependent transmission through the multipass amplifier normalized to 1 for vanishing tilts.  
  Similarly, the blue curve is the tilt dependent part of $G_\mathrm{8-pass}^\mathrm{M2-45^\circ-pair}$ ($e^{-2.98\cdot {10}^7
    \phi ^2}$) for the same amplifier but M2 replaced by a retro-reflector.  }
\label{fig:transmission-simulation}
\end{figure}

As explained in Sec.~\ref{sec:method} an alternative way to quantify the sensitivity to misalignment effects is to measure the gain decrease of the multipass amplifier as a function of the disk tilt $\phi$.
The total gain through the amplifier can be readily quantified using Eqs.~(\ref{eq:transmission-1}) to (\ref{eq:transmission-3}) and assuming that there are no other losses than already included in the aperture transmission of the disk.
The total gain in this case takes the form
\begin{eqnarray}
  G_\mathrm{8-pass}^\mathrm{M2-single}\approx (G_0)^8 \Big(\frac{W^2}{w_\mathrm{in}^2 + W^2}\Big)^8\, e^{-1.07\cdot {10}^8 \phi^2 }
\label{eq:gain-amp-1}
\end{eqnarray}
for the 8-pass amplifier with simple M2 mirror, and
\begin{eqnarray}
  G_\mathrm{8-pass}^\mathrm{M2-45^\circ-pair}\approx (G_0)^8 \Big(\frac{W^2}{w_\mathrm{in}^2 + W^2}\Big)^8\, e^{-2.98\cdot {10}^7 \phi ^2}
\label{eq:gain-amp-2}
\end{eqnarray}
for the amplifier where M2 is replaced by a retro-reflector.
In these equations $G_0$ is the gain per pass (averaged over the transverse beam profile) while the approximate symbol accounts for the small variations of the beam sizes $w_\mathrm{in}$ at the various passes and the assumption that the gain does not decrease with the number of passes (no saturation).
The exponential terms in these equations encompass the losses caused by the beam excursion from the optical axis.
A plot of the two functions $G_\mathrm{8-pass}^\mathrm{M2-single}$ and $G_\mathrm{8-pass}^\mathrm{M2-45^\circ-pair}$ is shown in Fig.~\ref{fig:transmission-simulation} in red and blue, respectively.

\begin{figure}[bt]
\centering
\includegraphics[width= 0.9 \linewidth]{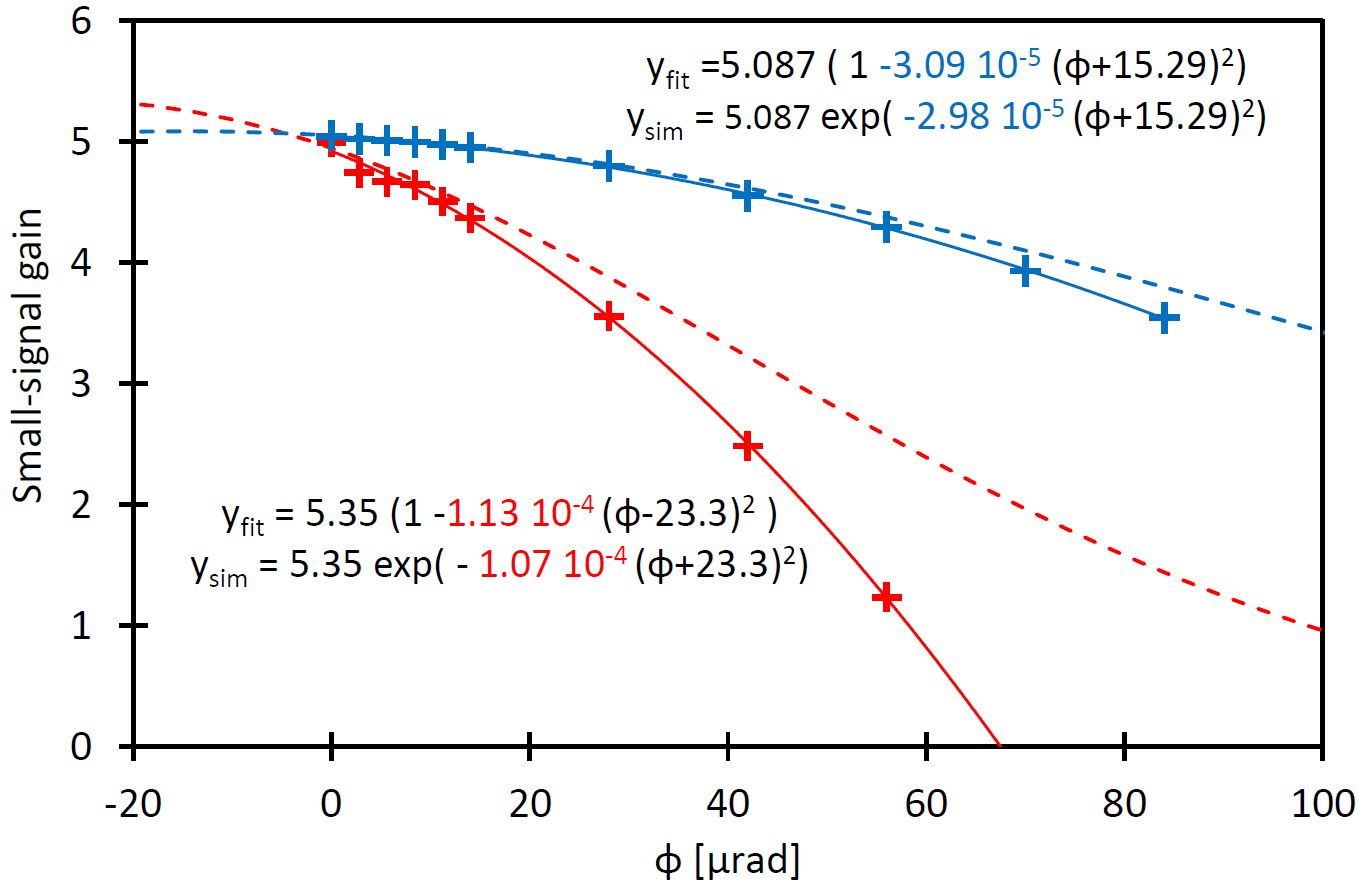}
\caption{
Small signal gain versus disk tilt in vertical direction.
The red points where measured for the multipass amplifier of Fig.~\ref{fig:scheme-old-amplifier} with simple M2-mirror.
The blue points have been obtained from the same amplifier but M2 replaced by a vertical retro-reflector. The measurements have been fitted with parabolic functions (solid lines). 
The dashed curves represent the predictions from Eqs.~(\ref{eq:gain-amp-1}) and (\ref{eq:gain-amp-2}) normalized to match the maxima of the parabolic fits. }
\label{image7}
\end{figure}

A measurement of the total gain versus disk tilt for the two multipass amplifier configurations (M2 simple, M2 as retro-reflector) is shown in Fig.~\ref{image7}.
The measured data were already presented in \cite{Schuhmann:15} but here they are compared with the theoretical predictions of Eqs.~(\ref{eq:gain-amp-1}) and (\ref{eq:gain-amp-2}).
It turned out that the best empirical fit functions to the measured data are given by parabola  and not by  Gaussian functions  as predicted in Eqs.~(\ref{eq:gain-amp-1}) and (\ref{eq:gain-amp-2}).

In our measurement the multipass amplifiers were not perfectly aligned so that the maximal gains were observed for non-vanishing disk tilts ($\phi=-23.3~\mu$rad and $\phi= -15.3~
\mu$rad extrapolated using the parabolic fits).
%
%the unperturbed situation ($\phi= 0$),
%
To compare the measurements to the model, the value of $(G_0)^8
(\frac{W^2}{w_\mathrm{in}^2 + W^2})^8$ of Eqs.~(\ref{eq:gain-amp-1}) and (\ref{eq:gain-amp-2}) has been fixed to the gain maximum obtained from the parabolic fit.
Also the angle $\phi$ in Eqs.~(\ref{eq:gain-amp-1}) and (\ref{eq:gain-amp-2}) has been redefined to account for the offset of the maximum (see parameterizations given in Fig.~\ref{image7}).
The resulting functions, shown by the dashed curves in Fig.~\ref{image7} that assume $W=4 w_\mathrm{in}$, are not fit to the data, but simply normalized to the maximum obtained from the parabolic fits.
For small excursions, there is an excellent agreement between the models expressed by Eqs.~(\ref{eq:gain-amp-1}) and (\ref{eq:gain-amp-2}) on the one hand and the measurements of Fig.~\ref{image7} on the other hand.
This agreement quantitatively confirm our assumption that the effective losses at the active medium can be described by a Gaussian aperture with $W \approx 4w_\mathrm{in}$ when $w_\mathrm{in} \approx 0.7R_p$.

For large tilts the measured losses exceed the theoretical predictions.
Indeed, the Gaussian approximation of the  aperture becomes increasingly inadequate as the beam axis approaches the edge of the super-Gaussian aperture.
However,  only small misalignments -- which are well described by our model -- are relevant  for the evaluation of a laser system.  
Lasers with exceeding  fluctuations and drifts of the output power  are in fact typically unsuitable for the practical applications. 

\section{Reduction of alignment sensitivity using an active compensation}

A significantly higher stability to misalignment can be realized by implementing a system actively controlling the alignment (tilt) of mirrors.
Auto-alignment units comprising quadrant detectors (or other devices that measure the beam position) feedback loops, and motorized mirror holders are commercially available and simple to implement~\cite{Aligna}.
However, obtaining the best possible alignment using the smallest number of controls requires the correct placement of quadrant detectors and active mirrors.

In this section, a simple  active stabilization system for the Fourier-based multipass amplifier of Fig.~\ref{fig:scheme-old-amplifier} is presented that compensates for tilts of the in-coupled beam, disk, M1 and M2 to overall minimize beam excursions.
The attention has been focused primarily to correct for tilts of the disk, as it is prone to misalignments owing to the pumping and cooling processes that it undergoes and because of the air wedge at its surface \cite{dietrich2017passive}.
It is assumed that the disk itself cannot be equipped with actuators to control its tilt.

\begin{figure}[tb]
\centering
\includegraphics[width= .9 \linewidth]{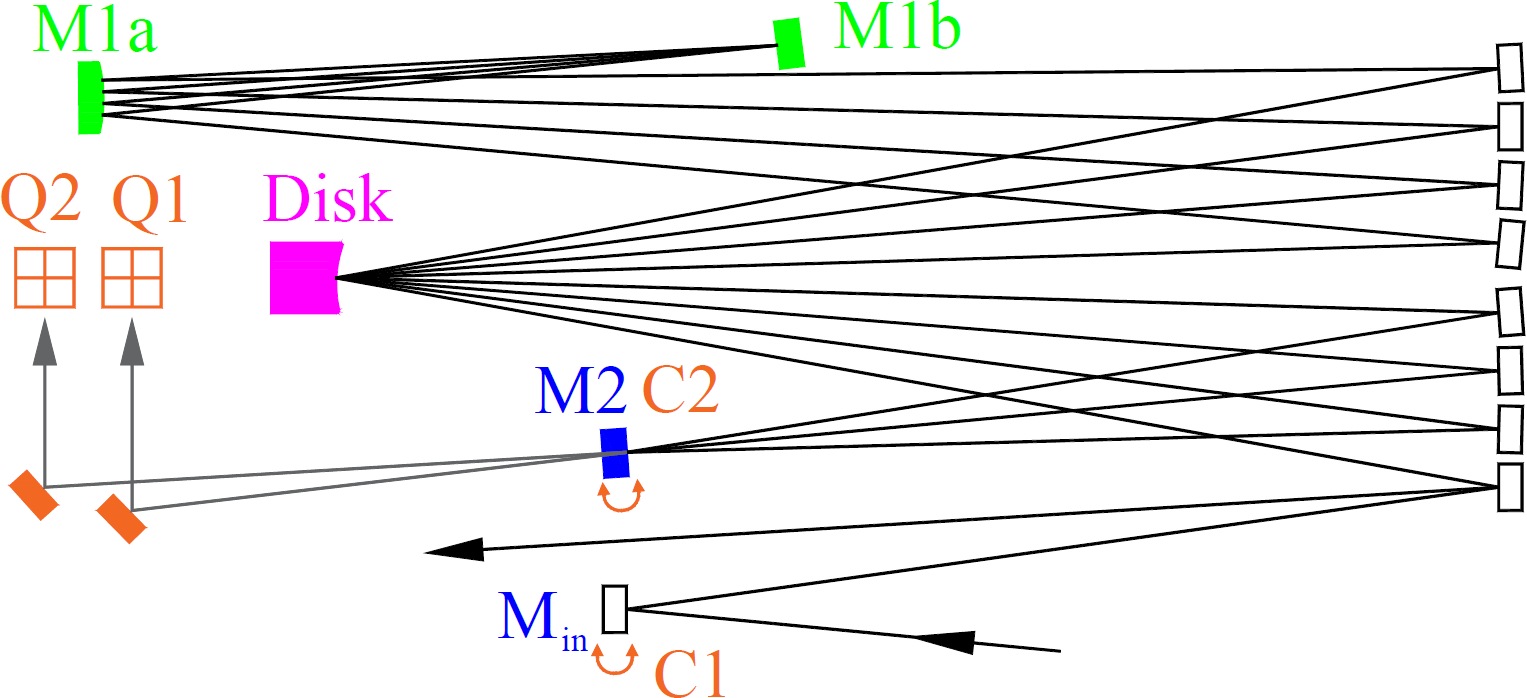}
\caption{Schematic of the realized multipass amplifier equipped with
  a simple auto-alignment system.  Only two loops (each with a vertical
  and a horizontal degree of freedom) are sufficient to mitigate the
  excursion of the laser beam from the optical axis for tilts of the
  disk, M1a, M1b, M2 and input beam. Each loop (C1 and C2)
  comprises a quadrant detector (Q1 and Q2) whose error signals act on
  the  motorized mirrors   M$_\mathrm{in}$ and M2, respectively.
  Ideally,  the quadrants measure the position of the beam
  at the first and second pass on the M2 mirror.  }
\label{image9}
\end{figure}

Figure~\ref{image9} shows the multipass amplifier of Fig.~\ref{fig:scheme-old-amplifier} upgraded with an optimal and simple active stabilization system.
The input beam direction is actively controlled by the feedback loop (C1) acting on the actuators of mirror M$_\mathrm{in}$ with two independent degrees of freedom (vertical and horizontal).
The beam travels successively to the disk where it undergoes the first amplification and reflection.
For the disk tilted (misaligned) by an angle $\phi$, the beam acquires an additional angle w.r.t. the optical axis of $\Delta \theta=2\phi$.
The Fourier propagation that takes place in the propagation disk-M1b-disk transforms the tilt of the beam leaving the disk into a beam excursion when the beam returns for the second time (second pass) to the disk.
At the disk, the beam suffers an additional tilt $\Delta \theta$.
Because of the short propagation length between disk and M2, the excursion of the beam at the mirror M2 is dominated by the tilt of the beam after the first reflection at the disk that is given by the tilt of the disk and the tilt of the beam before impinging for the first time on the disk.
The excursion of the beam at M2 can be measured by placing a quadrant detector (Q1) in the vicinity of M2 that measures the beam spuriously transmitted through M2.
Any deviation of the beam position from the set point (when the amplifier is aligned) generates two error signals: one for the horizontal, the other for the vertical direction.
Through the feedback loop C1, these error signals act on the actuators of mirror M$_\mathrm{in}$ to cancel the beam excursion at M2.

\begin{figure}[tb]
\centering
\includegraphics[width= \linewidth]{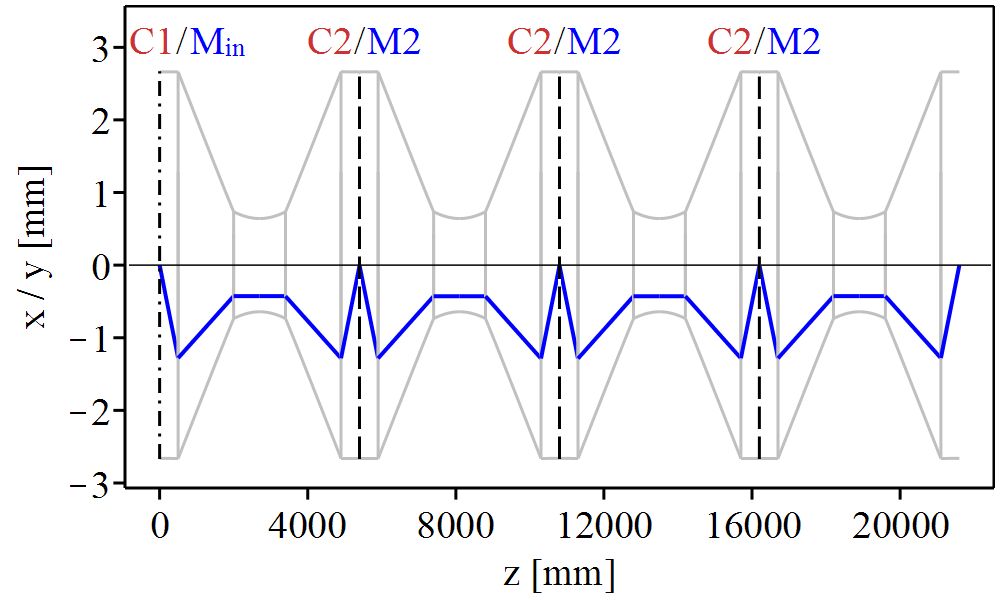}
\caption{
Similar to Fig.~\ref{fig:tilt-propagation-1} but for a multipass amplifier equipped with the simple auto-alignment system depicted in Fig.~\ref{image9}. 
Because of the efficient compensation produced by the active stabilization, the disk tilt has been increased to $\phi=1.25$~mrad, (a factor 25 times larger than in Fig.~\ref{fig:tilt-propagation-1}).
The correction generated by C1 (steering of M$_\mathrm{in}$) is compensating the downstream tilts occurring in the first and second pass on the disk, so that the beam excursion of the first pass on M2 is nullified.  
Similarly, the second loop C2 adjusts the tilt of the mirror M2 so that the beam excursions at all successive passes (third, forth etc.) on M2 are nullified.}
\label{image10}
\end{figure}

After reflection on mirror M2, the beam reaches the disk for the third time where it again acquires an angle $\theta=2\phi$.
This tilt once again results in an excursion from the set point position when the beam  reaches M2 for the second time.
The excursion of the second pass on M2 can be measured with a second quadrant detector (Q2).
The error signals generated by Q2 are used to regulate the actuators of the mirror M2 through a second feedback loop C2 so that the beam excursion at the second pass on mirror M2 is nullified.
%
%All the following propagation from mirror M2 back to mirror M2 are exposed to the same mirror tilt. 
%
%Therefore, all successive passes are corrected for by the tilt of the mirror M2.
%
The same correction (tilt) of mirror M2 automatically correct all the successive passes on M2 due to the
repetitive structure of the amplifier (see Fig.~\ref{image10}).
Both loops are sufficiently independent so that their interplay can compensate for tilts of the in-coupled beam and for tilts of the disk.  
The regulation can also partially compensate for tilts of M1.

Only two feedback loops are therefore sufficient to stabilize the most critical optical elements of the multipass amplifier for all passes as M1, M2, and active medium are common for all passes.
Oppositely, it has to be stressed that individual mirror misalignments within the mirror array cannot be compensated by this feedback system so that high stability is required for the individual holders of the array mirrors.

Both quadrants have to be placed in the vicinity of mirror M2.
It is possible to detect the position of the first pass on M2 with Q1 and of the second pass on M2 with Q2 because the two passes impinge on M2 at different angles.
Thus, a short free propagation from M2 to the quadrants Q1 and Q2 can be used to separate the two spuriously transmitted beams (see Fig.~\ref{image9}).

In Fig.~\ref{image10}, the excursion of the laser beam axis from the optical axis is illustrated along the 8-pass amplifier of Fig.~\ref{image9} equipped with the above-described active stabilization.
Only the propagations for a tilted disk is displayed.
The propagation for a tilted in-coupled beam is not shown, as a tilt of the input beam can be exactly canceled by the loop C1 so that the beam would propagate on the optical axis ($x=y=0$) of the amplifier.
The resulting excursion evolution shown in Fig.~\ref{image10}  can be compared with the excursions in Figs.~\ref{fig:tilt-propagation-1} and \ref{fig:tilt-propagation-2}.
Similar overall excursions are visible but in Fig.~\ref{image10} the disk tilt has been increased by a factor of 25 compared to the non-active stabilized amplifiers.
Hence, the active stabilization improves greatly the stability to disk tilts.

\begin{figure}[tb]
\centering
\includegraphics[width= \linewidth]{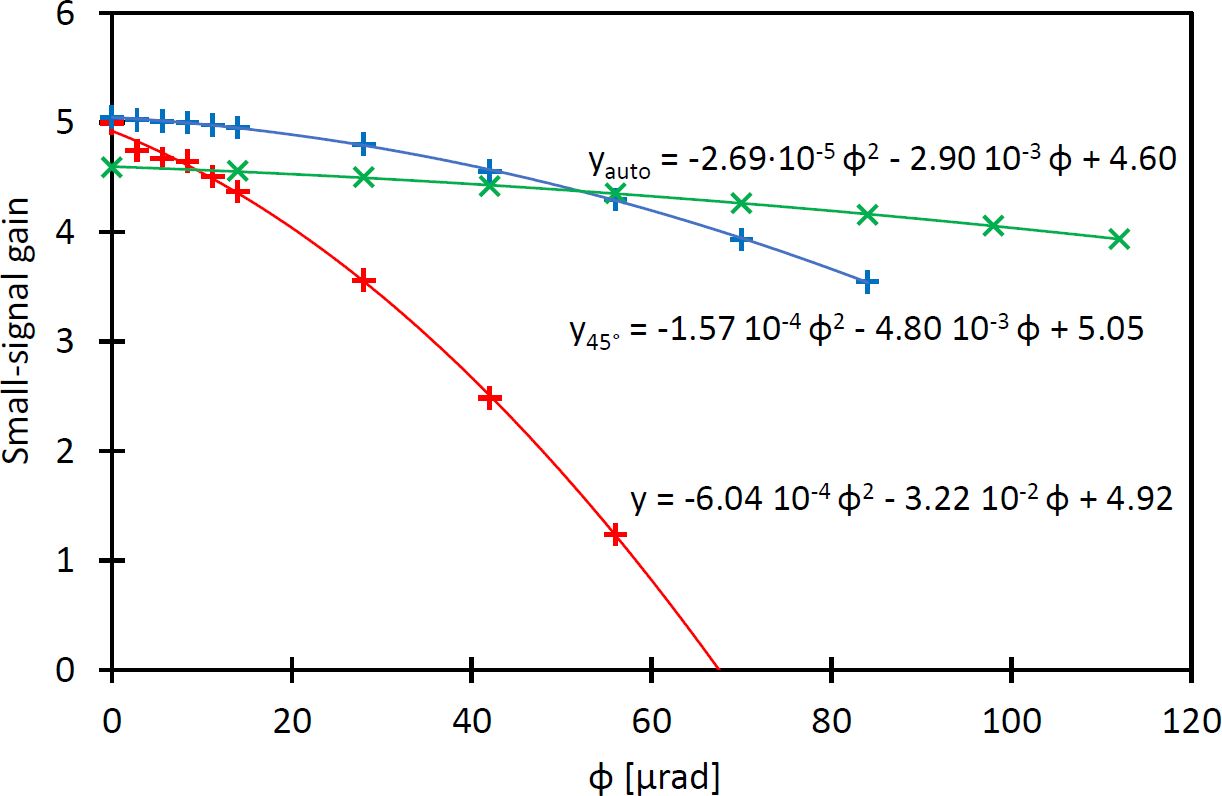}
\caption{Measured small-signal gain for three 8-pass amplifiers versus the disk tilt $\phi$. 
The red points have been taken for the simple amplifier of Fig.~\ref{fig:scheme-old-amplifier}, the blue points for the same amplifier but M2 replaced by a vertical retro-reflector, the green ones for the amplifier of Fig.~\ref{image9} equipped with the active stabilization system.
Polynomial fits have been drawn to guide the eye. 
For the amplifier with the active stabilization system the mirror M2 has been replaced with a mirror with slightly higher transmission to generate a robust error signal from Q1 and Q2. 
This reduces the overall gain of the amplifier but does not alter its tilt dependence.}
\label{image11}
\end{figure}

The auto-alignment system of Fig.~\ref{image9} has been tested experimentally.
The measured gain decrease of the actively controlled amplifier for variations of the disk tilt is summarized in Fig.~\ref{image11} and compared with the results from the non-actively stabilized amplifiers with the same optical layout.
The actively stabilized amplifier shows a major (order of magnitude) decrease of the sensitivity to disk tilts.
However, the measured stability improvement was smaller than predicted from simulations.
This reduced performance can be ascribed to the non-vanishing M2-Q1 and M2-Q2 distances that in the realized amplifier were about 1.5~m.
This issue could be solved by implementing an imaging of M2 on the two quadrants.

It is interesting to note that the number of actively controlled mirrors can be reduced to one without loss of alignment stability by eliminating the loop C2 and replacing  the active mirror M2 with a corner-cube reflector with 3 mirrors arranged at an angle of $54^\circ$ and acting as retro-reflectors. 

\section{Summary}

A simple model has been presented to calculate the losses occurring at the soft aperture naturally present in a pumped active medium and the effects of this aperture for the beam propagation.
This knowledge has been used to quantify the misalignment sensitivity of multipass amplifiers in two ways.
The first, by simply tracking the evolution of the beam excursion w.r.t. the optical axis while the beam propagates in the amplifier for a tilted (misaligned)  optical element or  input beam.
The second, which requires the knowledge of the first, by computing the decrease of the amplifier gain as a function of the tilt of the considered optical element.

These two methods have been used to investigate the sensitivity of Fourier-based amplifiers to beam tilts.
The choice of the Fourier-based amplifiers is motivated by the superior stability of these multipass amplifiers to variations of thermal lens and aperture effects at the active medium compared with state-of-the-art amplifiers based on the 4f-imaging propagation.
An extensive comparison between the two multipass architectures can be found in ~\cite{Schuhmann:18a}.
Three variations of the same Fourier-based amplifier were investigated in this study: the basic layout of Fig.~\ref{fig:scheme-old-amplifier}, the basic layout but mirror M2 replaced with a vertical retro-reflector, and the basic layout equipped with an auto-alignment system shown Fig.~\ref{image9}.

The simulations were compared with measurements of the amplifier gain and good agreement has been found.
The stability of Fourier-based multipass amplifiers has been demonstrated via months long operation of a thin-disk laser based on the layout of Fig.~\ref{fig:scheme-old-amplifier}  without the need of realignment.
Furthermore, simulations and observations showed that this stability can be improved by a factor of 4 in one direction (vertical) when M2 is implemented as a pair of $45^\circ$ mirrors acting has vertical retro-reflectors, or in both directions (vertical and horizontal) when M2 is implemented as a three-mirror system acting as retro-reflectors in both directions.
Alike, more than an order of magnitude improvement in tilt stability can be obtained by equipping the multipass amplifier with a simple auto-alignment system comprising only two quadrant detectors and two motorized mirror holders.

These findings combined with the investigation presented in~\cite{Schuhmann:18a} fully qualifies the Fourier-based design to be the appropriate choice in the high energy and high power sector.
\section{Funding information}
We acknowledge the support from the Swiss National Science Foundation Project SNF 200021\_165854, the European Research Council ERC CoG. \#725039 and ERC StG. \#279765. The study has been also supported by the ETH Femtosecond and Attosecond Science and Technology (ETH-FAST) initiative as part of the NCCR MUST program.

\bibliographystyle{unsrt}   
\bibliography{ms}

\end{document}